\documentclass[12pt]{article} 
\usepackage{epsfig}
\usepackage{axodraw}
\usepackage{a4}
\usepackage{latexsym}
\usepackage{cite}

\usepackage{pslatex}
\usepackage[latin1]{inputenc}
\usepackage[T1]{fontenc}

\usepackage{color,colordvi}
\def\BLUE#1{\Blue{#1}}

\newcommand{\eqmel}{\raisebox{-0.07cm}{$\:\stackrel{{\rm M}}{=}\:$} }

\newcommand{\hspn}{{\hspace{-4mm}}}
\newcommand{\beq}{\begin{equation}}
\newcommand{\eeq}{\end{equation}}
\newcommand{\bea}{\begin{eqnarray}}
\newcommand{\eea}{\end{eqnarray}}
\newcommand{\nn}{\nonumber}
\newcommand{\MSb}{$\overline{\mbox{MS}}$}
\newcommand{\as}{\alpha_{\rm s}}
\newcommand{\ar}{a_{\rm s}}
\newcommand{\ra}{\rightarrow}
\newcommand{\ep}{\epsilon}
\newcommand{\GE}{\gamma_{\rm e}}

\begin{document}

\setlength{\parskip}{0.2cm}
\setlength{\baselineskip}{0.53cm}

\def\Ftwo{{F_{\:\! 2}}}
\def\Qs{{Q^{\, 2}}}
\def\ca{{C_A}}
\def\cas{{C^{\: 2}_A}}
\def\cf{{C_F}}
\def\nf{{n^{}_{\! f}}}

\def\caf{{C_{A\!F}}}
\def\cafs{{C^{\,2}_{\!A\!F}}}
\def\caft{{C^{\,3}_{\!A\!F}}}
\def\cafn{{C^{\:n}_{\!A\!F}}}
\def\cafnmo{{C^{\:n-1}_{\!A\!F}}}

\def\x1{{(1 \! - \! x)}}

\begin{titlepage}

\noindent
\hspace*{\fill} LTH 875\\[1mm]
\hspace*{\fill} May 2010 \\
\vspace{2.5cm}
\begin{center}
\Large
{\bf Leading logarithmic large-$x$ resummation of off-diagonal\\[1mm]
splitting functions and coefficient functions}\\
\vspace{2cm}
\large
A. Vogt\\
\vspace{1cm}
\normalsize
{\it Department of Mathematical Sciences, University of Liverpool \\
\vspace{1mm}
Liverpool L69 3BX, United Kingdom}\\[2.5cm]
\vfill
\large
{\bf Abstract}
\vspace{-0.2cm}
\end{center}
We analyze the iterative structure of unfactorized partonic structure 
functions in the large-$x$ limit, and derive all-order expressions for the
leading-logarithmic off-diagonal splitting functions $P_{\rm gq}$ and 
$P_{\rm qg}$ and the corresponding coefficient functions $C_{\phi,\rm q}$ and 
$C_{2,\rm g}$ in Higgs- and gauge-boson exchange deep-inelastic scattering. 
The splitting functions are given in terms of a new function not encountered in
perturbative QCD so far, and vanish maximally in the supersymmetric limit 
$\,\ca\! - \cf \!\ra\, 0$. The coefficient functions do not vanish in this 
limit, and are given by simple expressions in terms of the above new function 
and the well-known leading-logarithmic threshold exponential.
Our results also apply to the evolution of parton fragmentation functions and 
semi-inclusive $e^{+} e^{-}$ annihilation.
   
\vspace{1cm}
\end{titlepage}
%
%
\noindent
The splitting functions $P_{\,ik}(x,\as)$, $\,i,\:k \,=\, \rm q,\: g$,  
governing the scale dependence of the light-quark and gluon distributions of 
hadrons are among the most important quantities in perturbative QCD. In the 
helicity-averaged case these universal, but factorization-scheme dependent 
functions are completely known to the third order in the strong coupling 
constant $\as$ \cite{OPE-LO,AP77,OPE-NLO,CFP80,MVV34}. 
Those results, as well as the computation of the second Mellin moment of the 
(flavour non-singlet) quark-quark splitting function to order $\as^{\,4}$ 
\cite{BC-LL06}, show a perturbative expansion which is remarkably well-behaved 
away from the momentum-fraction endpoints $x=0\,$ and $x=1$.  Without loss of 
information identifying the renormalization and mass-factorization scales, we 
write this expansion as
\beq
\label{Pexp}
  P_{\,ik}(x,\as) \:\: = \:\: \sum_{n=0}^\infty \ar^{\,n+1}\, P^{\,(n)}_{ik}(x)
  \quad \mbox{with} \quad \ar \;\equiv\; {\as \over 4\pi} \;\; .
\eeq
In the small-$x$ (high-energy) limit all four flavour-singlet splitting 
functions exhibit a single-log\-arithmic higher-order enhancement, i.e., terms 
of the form $\as^{\,n+1}\, x^{\,-1} \ln^{\,n-a} x\,$ occur at (almost) all 
orders with $a \geq a_{\rm min}= 0\,$ for $P_{\rm gg}$ and $P_{\rm gq}$, and 
$a_{\rm min}= 1$ for $P_{\rm qq}$ and $P_{\rm qg}$ \cite{BFKL}.
The contributions for $a = 1$ have been obtained \cite{CH94,NL-BFKL} except for 
$P_{\rm qg}$. Consequently only $P_{\rm gg}$ is known at next-to-leading 
logarithmic (NLL) small-$x$ accuracy at this point. 

In this letter we address the large-$x$ (soft-gluon) limit. It is useful to 
switch to Mellin moments,
\beq
\label{Mtrf}
 f(N) \;=\; \int_0^1 \! dx \, 
 \left(\, x^{\,N-1} \{ - 1 \} \right) \: f(x)_{\{+\}}
 \:\: ,
\eeq
where the parts in curly brackets refer to the case of $(1-x)^{-1}$ 
$+$-distributions. 
Keeping only the leading -- and subleading, if $\ln^{\,k} N$ is replaced by 
$\ln^{\,k} N + k\,\GE \ln^{\,k-1} \!N$ at any stage -- contributions, the 
relations between the relevant expressions in $x$-space and Mellin-$N$ space 
are given~by
\beq
\label{Logtrf}
  \frac{\ln^{\,n} \!\x1}{\x1_+} \;\:\eqmel\;\:
  \frac{(-1)^{n+1}}{n+1} \, \ln^{\,n+1} \!N \:+ \:\ldots
\:\: , \quad
  \ln^{\,n} \!\x1  \;\:\eqmel\;\: 
  \frac{(-1)^n}{N} \, \ln^{\,n} \! N  \: + \:\ldots
\eeq
with $\eqmel$ denoting equality under the Mellin transformation~(\ref{Mtrf}).

The dominant and subdominant ($\,N^{\,0}$ and $N^{\,-1\,}$) large-$x$ 
contributions to the diagonal splitting functions $P_{\,\rm qq}$ and 
$P_{\,\rm gg}$ in Eq.~(\ref{Pexp}) are stable in the usual \MSb\ factorization 
scheme \cite{BBDM78}, i.e., their form
\beq
\label{Pdiag}
  P_{\rm qq/gg}^{\:(n-1)}(N) \;\: = \:\: \mbox{}
     -\: A^{\,(n)}_{\,\rm q/g} \;\ln \,N  
   \:+\: B^{\,(n)}_{\rm q/g} 
   \:-\: C^{\,(n)}_{\rm q/g} \; N^{\, -1}\ln \,N
   \:+\: \ldots
\eeq
is the same at all orders $n$ \cite{Korch89,DMS05}. 
The quark and gluon cusp anomalous dimensions are related by $A_{\rm g} = 
\ca/\cf\; A_{\rm q}$ \cite{Korch89}, and the coefficients $C^{\,(n)}$ are 
functions of lower-order quantities $A^{(k)}$ \cite{MVV34,DMS05}. 
The $1/N$-suppressed off-diagonal splitting functions $P_{\rm gq}$ and 
$P_{\rm qg}$, on the other hand, include a double-logarithmic higher-order 
enhancement with a particular colour structure,
\bea
\label{Poffd}
  C_F^{\,-1} P_{\rm gq}^{\,(n)}  \: = \;
  \nf^{\!\!-1} P_{\rm qg}^{\,(n)\!} 
  &\!\!\! =\! &  
  N^{\,-1} \ln^{\,2n} \!N \:\:D_0^{\,(n)} \,\cafn
\nn \\[0.5mm] & & \mbox{\hspn}\!\! +
  N^{\,-1} \ln^{\,2n-1} \!N \, 
    \left[ D_{1,\,AF}^{\,(n)}\,\caf + \,D_{1,F}^{\,(n)}\,\cf 
    + \,D_{1,T}^{\,(n)} \,\nf \right] \, \cafnmo
  + \,\ldots \;\; . \quad
\eea
Here $\ca$ and $\cf$ are the usual SU(N$_c$) colour factors, which $\ca = 
{\rm N}_c = 3$ and $\cf = 4/3$ in QCD. $\caf \,\equiv\, \ca - \cf$, and $\nf$ 
stands for the number of light flavours. 
All double logarithmic terms, $\,\ln^{\,k} \!N$ with $\,n+1 \leq k \leq 2n\,$ 
vanish for $\cf = \ca$, which is part of the colour-factor choice leading to an
${\cal N}\! =\! 1$ supersymmetric theory. The leading coefficients 
$D_0^{\,(n)}$ -- which have the same modulus for $P_{\rm gq}$ and 
$P_{\rm qg\,}$ --
vanish maximally, i.e., with the highest possible power of $\caf$, the 
next-to-leading contributions $D_1$ -- which are not same for $P_{\rm gq}$
and $P_{\rm qg}$ -- next-to-maximally etc. 
These properties and the coefficients $D_{\!j}$ are known from the diagram 
calculations in Refs.~\cite{CFP80,MVV34} to order $\as^{\,3}$.
  
Eq.~(\ref{Poffd}) and the determination of the coefficients $D_0$, $D_1$ 
and $D_2$ has been extended to order $\as^{\,4}$ in Ref.~\cite{SMVV1}. Those 
results have been deduced from the -- formally yet unproven -- single-%
logarithmic large-$x$ behaviour of the physical evolution kernels $K(N,\as)$
for the system $(F_2, F_\phi)$ of flavour-singlet gauge-boson and Higgs 
exchange (in the heavy top-quark limit, see also Ref.~\cite{DGGLcphi}) 
structure functions in deep-inelastic scattering (DIS) \cite{FP82}, 
\beq
\label{Kdef}
  \frac{d F}{d \ln \Qs} \:\: =\:\:
  \Big( \, \beta(\ar)\: \frac{d\, C}{d \ar}_{\,\!}
  + C\, P\, \Big)\, C^{\,-1}\, F \;\: \equiv \;\: K F
\eeq
with
\beq
\label{FCPKsg}
  F \:=\:
  \Big( \begin{array}{c} \!\! \Ftwo \!\! \\ \! F_\phi \!\!
        \end{array} \Big)
\; , \;\;\;
  C \:=\:
  \Big( \! \begin{array}{cc} C_{2,\rm q}^{}\! & C_{2,\rm g}^{}\! \\
  C_{\phi,\rm q}\! & C_{\phi,\rm g}\! \end{array}\! \Big)
\; , \;\;\;
 P \:=\:
 \Big( \! \begin{array}{cc} P_{\,\rm qq} \! & P_{\,\rm qg} \! \\
  P_{\,\rm gq}\! & P_{\,\rm gg}\! \end{array}\! \Big)
\; , \;\;\;
 K \:=\:
 \Big( \! \begin{array}{cc} K_{22}\! & K_{2\phi}\! \\
  K_{\phi 2}\! & K_{\phi\phi}\! \end{array}\! \Big)
 \:\: ,
\eeq
in conjunction with the three-loop coefficient functions computed in 
Refs.~\cite{MVV6,SMVV1}. 
In particular it turned out that the fourth-order coefficient $D_0^{\,(3)}$ 
vanishes, a fact that was attributed to an accidental cancellation of 
contributions. The single-log enhancement of the physical kernels provides 
relations between double-logarithmic contributions to the singlet splitting 
functions and coefficient functions also beyond this order but, unlike in
corresponding non-singlet cases \cite{MV35} which include Eq.~(\ref{Pdiag}) but
not Eq.~(\ref{Poffd}), no definite higher-order predictions of any expansion 
coefficients.

In the present letter we derive an all-order expression for the $\as^{\,n+1\!} 
\ln^{\,2n} (1-x)$ leading-logarith\-mic (LL) large-$x$ contributions 
to the off-diagonal splitting functions $P_{\rm gq}$ and $P_{\rm qg}$.
This derivation is based on the large-$x$ properties of the unfactorized 
expressions for the respective gluon and quark contributions to the structure 
functions $\Ftwo$ and $F_\phi$ in dimensional regularization. Hence we obtain 
the corresponding $\as^{\,n} \ln^{\,2n-1} (1-x)$ contributions to the 
off-diagonal coefficient functions $C_{2,\rm g}$ and $C_{\phi,\rm q}$ in 
Eq.~(\ref{FCPKsg}) as well. 
Our results also answer the questions whether or not $D_0^{\,(3)}=0$ in 
Eq.~(\ref{Poffd}) is really accidental (it is not), and whether or not at least
the leading double-logarithmic large-$x$ contributions to Eq.~(\ref{Kdef}) 
definitely vanish at all orders in $\as$ (they do). 

The above-mentioned feature of the physical evolution kernel suggests an
iterative structure of the unfactorized partonic structure functions or 
forward Compton amplitudes. For brevity suppressing, as already done in 
Eqs.~(\ref{Kdef}) and (\ref{FCPKsg}) above, all functional dependences on $N$, 
$\as$ and the dimensional offset $\ep$ with $D = 4 - 2\ep$, these quantities
can be factorized as (cf., e.g., Refs.~\cite{MVV6,SMVV1})
\beq
\label{Tdec}
  T_{a, k} \;\;=\;\; \widetilde{C}_{a, i} \: Z_{\,ik}
\:\: .
\eeq
Here the (process-dependent) $D$-dimensional coefficient functions 
$\widetilde{C}_{a, i}$ consists of contributions with all non-negative powers 
of $\ep$. The universal transition functions (or, in the language of the 
operator-product expansion (OPE),  
renormalization constants) $Z_{\,ik}$ collecting all negative powers of $\ep$ 
are related to the splitting functions in Eq.~(\ref{FCPKsg}) (or the anomalous 
dimension $\gamma$ of the OPE) by
\beq
\label{PofZ}
  - \:\!\gamma \;\:=\;\: P \;\:=\;\:
 \frac{d\:\! Z }{d\ln \Qs }\: Z^{-1}
\:\: ,
\eeq
where we have, again without losing any information, identified the 
renormalization and factorization scale with the physical hard scale $\Qs$. 
Using the $D$-dimensional evolution of the coupling,
\beq
\label{arunD}
  \frac{d\:\! \ar}{d\ln \Qs} \;\:=\;\: -\,\ep\, \ar + \beta (\ar)
\eeq
where $\beta (\ar)$ denotes the usual four-dimensional beta function of QCD,
$\beta (\ar) = - \beta_0 \,\ar^2 + \,\ldots$ with 
$\beta_0 = 11/3\:\ca - 2/3\:\nf\,$, 
Eq.~(\ref{PofZ}) can be solved for $Z$ order by order in $\as$.
 
In general, the higher-order coefficients $Z^{(n)}$ become very complicated
at higher powers $n$ of~$\ar$. Here, however, we are interested only in the
LL contributions at order $N^{\,-1}$ for $Z_{\rm qg}$ and $Z_{\rm gq}$ and 
$N^{\,0}$ for $Z_{\rm qq}$ and $Z_{\rm gg}$ (required for Eq.~(\ref{Tdec}) also 
in the off-diagonal cases). Consequently there can be at most one off-diagonal 
$N^{\,-1}$ factor per term, and $P_{\rm qq,\,gg}^{(n \geq 1)}$ can be neglected at leading- and also next-to-leading logarithmic (NLL) accuracy due to
Eq.~(\ref{Pdiag}). Finally $\beta (\ar)$ in Eq.~(\ref{arunD}) only enters at 
the NLL level.
For the off-diagonal $i \neq k$ entries these simplifications lead to
\bea
\label{ZikLL}
  Z_{ik}^{\,(n)} &\! \cong\! &
  {1 \over n!}\: \ep^{-n} \;\sum_{l=0}^{n-1} 
    \left( \gamma^{\,(0)}_{\,ii} \right)^{n-l-1} \gamma^{\,(0)}_{\,ik}
    \left( \gamma^{\,(0)}_{\,kk} \right)^l
\nn \\[1mm] & & \mbox{\hspn} \! + \,
  {1 \over n!} \;\sum_{m=1}^{n-1} \ep^{-n+m} \;\sum_{l=0}^{n-m-1}\; 
    {(m+l)! \over l!}
    \left( \gamma^{\,(0)}_{\,ii} \right)^{n-m-l-1} \gamma^{\,(m)}_{\,ik}
    \left( \gamma^{\,(0)}_{\,kk} \right)^l 
\eea
with, always keeping the LL contributions only, 
$\,\gamma^{\,(0)}_{\,\rm qq} =   4\, \cf \ln N$, 
$\,\gamma^{\,(0)}_{\,\rm gg} =   4\, \ca \ln N$,
$\,\gamma^{\,(0)}_{\,\rm qg} = - 2   \nf\, N^{\,-1}$ and
$\,\gamma^{\,(0)}_{\,\rm gq} = - 2\, \cf\, N^{\,-1}$.
The corresponding diagonal quantities $Z_{ii}^{\,(n)\!}$, 
$\,i, \,=\, \rm q,\: g$, are simply given by
\beq
\label{ZiiLL}
  Z_{ii}^{\,(n)} \;\; \cong \;\;
  {1 \over n!}\: \ep^{-n} \left( \gamma^{\,(0)}_{\,ii} \right)^n
\:\: .
\eeq
Here and below $\,\cong\,$ denotes equality if NLL contributions on both sides 
are neglected.
 
The $\ep^{-n} \ldots \ep^{-2}$ contributions at the $n$-th order in $\ar$ of
the products (\ref{Tdec}) include only lower-order quantities and thus provide
a (at high $n$ large) number of consistency checks. The $\ep^{-1}$ and $\ep^0$
terms include the desired $n$-loop contributions to the splitting functions and 
(four-dimensional) coefficient functions, respectively, in Eq.~(\ref{FCPKsg}). 
In order to determine these quantities at a higher order $\ar^{\,l}$, also the
coefficients of $\ep^{\,k}$ with $ 0 < k \leq l-n$ are required. Hence an 
all-order determination of the splitting functions and coefficient functions 
requires expressions for $T_{a, k}$ which are, at the logarithmic accuracy
under consideration, exact in both $\as$ and $\ep$.

The first four $\ep^{-k}$ coefficients of the amplitudes $\,T_{\phi, 
\rm q}^{\,(n)}$ and $\,T_{2, \rm g}^{\,(n)\!}$ can be determined at all orders 
$n$ from the third-order calculations in Refs.~\cite{MVV34,MVV6,SMVV1} and the
all-order mass-factorization (or OPE) relation (\ref{Tdec}) with 
Eqs.~(\ref{ZikLL}) and (\ref{ZiiLL}). These results are of the form 
\beq
\label{TijLL}
  {1 \over \cf} T_{\phi, \rm q}^{\,(n)} \;\: \cong \;\:
  {1 \over \nf} T_{2, \rm g}^{\,(n)} \;\: \cong \;\:
  {\ln^{\,n-1} N \over N \ep^n} \sum_{m=0}^\infty (\ep \ln N)^m {\cal L}_{n,m} 
    \left( C_{\!F}^{\,n\!-\!1\!} + C_{\!F}^{\,n\!-\!2} \ca + \ldots 
         + C_{\!A}^{\,n-1} \right)
\:\: ,
\eeq
i.e., the leading-logarithmic expansion coefficients ${\cal L}_{\:\!n,m}$ at a
given order in $\ar$ and $\ep$ are the same for both off-diagonal amplitude and
all contributing colour factors. 
Eq.~(\ref{TijLL}) is the first ot two equations with a clear-cut all-$\ep$ 
structure to all orders in $\as$ which, unavoidably, is guaranteed only to a 
finite depth (here $m=3$) in $\ep$ by previous results. However, the simplicity
of the structure and the tight functional forms of the $D$-dimensional 
expressions, see Eqs.~(\ref{Tphiqall}) - (\ref{T2q1}) below, very strongly 
suggest that the all-$\ep$ form is indeed correct. 
Also an inspection of the ladder-type diagrams generating the leading 
logarithmic large-$x$ contributions to $T_{\phi, \rm q}$, illustrated in 
Fig.~1$\,$(a), indicates that the $\as^{\,n}\, C_F^{\,n-k}\, C_A^{\,k}\,$, 
$0 < k < n$, LL terms have the same coefficients as their 
$\as^{\,n} C_F^{\,n}$ counterparts: any differences between different colour
factors would be of a combinatorial nature, and thus be obvious from the known
first powers in $\ep$. 
The situation is analogous for the case of $T_{2, \rm g}^{\,(n)}$, see 
Fig.~1$\,$(b).
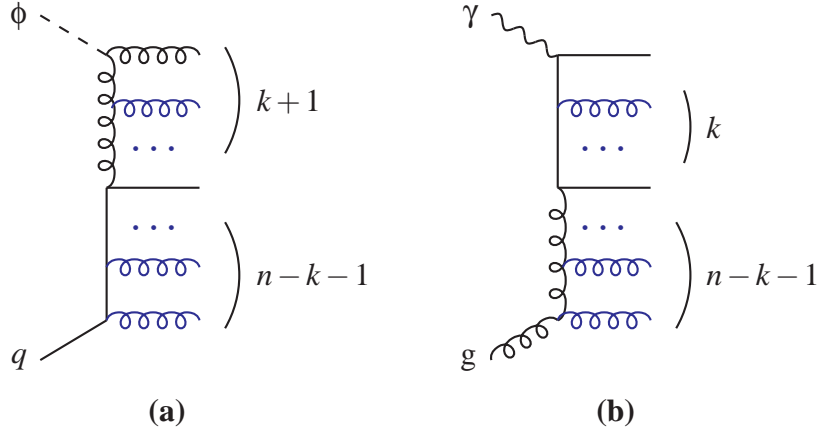
\begin{figure}[htb]
\vspace{-1mm}
\begin{center} 
\begin{picture}(280,160)(0,-30)
\SetWidth{0.8}
\Line(5,-15)(30,0)
\Line(30,0)(30,50)
\Gluon(30,50)(30,100){-3}{5}
\DashLine(5,115)(30,100){4}
\Line(30,50)(65,50)
\Gluon(30,100)(65,100){3}{4}
\SetColor{Blue}
\Gluon(30,0)(65,0){3}{4}
\Gluon(30,20)(65,20){3}{4}
\Text(57,35)[r]{\BLUE{\Large $\ldots$}} 
\Text(57,65)[r]{\BLUE{\Large $\ldots$}} 
\Gluon(32,80)(65,80){3}{4}
\SetColor{Black}
\Text(60,-35)[r]{\bf (a)}
\Text(0,-15)[r]{$q$}
\Text(0,115)[r]{$\phi$}
\CArc(40,17)(40,-30,30)
\Text(87,17)[lc]{$n-k-1$} 
\CArc(40,83)(40,-30,30)
\Text(87,83)[lc]{$k+1$} 
%
\SetWidth{0.8}
\Gluon(175,-15)(200,0){3}{3}
\Gluon(200,0)(200,50){-3}{4}
\Line(200,50)(200,100)
\Photon(175,115)(200,100){2}{3}
\Line(200,50)(235,50)
\Line(200,100)(235,100)
\SetColor{Blue}
\Gluon(200,0)(235,0){3}{4}
\Gluon(202,20)(235,20){3}{4}
\Text(227,35)[r]{\BLUE{\Large $\ldots$}}
\Text(227,65)[r]{\BLUE{\Large $\ldots$}}
\Gluon(200,80)(235,80){3}{4}
\SetColor{Black}
\Text(230,-35)[r]{\bf (b)}
\Text(170,-15)[r]{$\rm g$}
\Text(170,115)[r]{$\gamma$}
\CArc(210,17)(40,-30,30)
\Text(257,17)[lc]{$n-k-1$}
\CArc(210,73)(40,-20,20)
\Text(257,73)[lc]{$k$}
\end{picture}
\end{center}
\vspace*{-1mm}
\caption{
\label{fig1}
Typical diagrams for the leading-logarithmic large-$x$ terms of the $n$-th 
order quantities $T_{\phi, \rm q}^{\,(n)}$ (left) and $T_{2,\rm g}^{\,(n)}$ 
(right) in Eqs.~(\ref{Tdec}) and (\ref{TijLL}). 
Shown are $C_F^{\,n-k}\, C_A^{\:k}$ and $\nf\, C_A^{\,n-k-1}\, C_F^{\:k}$ 
contributions to the former and latter expressions, respectively.}
\vspace*{1mm}
\end{figure}

It is therefore sufficient to derive the complete LL expression for just one 
colour structure of one of the two quantities in Eq.~(\ref{TijLL}) to all 
orders $n$. 
For this we choose the abelian $C_F^{\,n}$ parts of $T_{\phi, \rm q}^{\,(n)}$,
as Fig.~1$\,$(a) for $k=0$ suggests a factorization in terms of 
$T_{\phi, \rm q}^{\,(1)}$ and $T_{2, \rm q}^{\,(n-1)}$ for these quantities.
Indeed, the third-order results of Ref.~\cite{SMVV1} imply
\beq
\label{Tphiqn}
  T_{\phi, \rm q}^{\,(n)} \Big|_{\,C_F^{\:n}} 
  \,\; \cong \;\;
  {1 \over n} \: T_{\phi, \rm q}^{\,(1)}\, T_{2, \rm q}^{\,(n-1)} 
  \,\; \cong \;\;
  {1 \over n!} \: T_{\phi, \rm q}^{\,(1)}\, 
  \left( T_{2, \rm q}^{\,(1)} \right)^{n-1}
\: .
\eeq
Here the second equality is due to 
\beq
\label{T2qn}
 T_{2, \rm q}^{\,(n)} \;\;\cong \;\; 
 {1 \over n!} \: \left( T_{2, \rm q}^{\,(1)} \right)^n
\eeq
which, in conjunction with Eq.~(\ref{T2q1}) below, is equivalent to the 
well-known leading-logarithmic threshold-exponentiation result \cite{C2LL}
\beq
\label{C2res}
 C_{2,\rm q} \;\; \cong \;\; \exp \left( 2 \ar\, \cf \ln^{\,2} N \right)
\:\: .
\eeq
Eq.~(\ref{Tphiqn}) is the second of the two all-$\ep$ relations mentioned below 
Eq.~(\ref{TijLL}), and the comments made there also apply here. 
Collecting the $\as$-expansion coefficients (\ref{Tphiqn}), one arrives at the 
closed all-order expression
\beq
\label{Tphiqall}
  T_{\phi, \rm q}\Big|_{C_{\!F}\: \rm only} \;\, \cong \;\;
  T_{\phi, \rm q}^{\,(1)} \:\: 
  { \exp \left( \ar T_{2, \rm q}^{\,(1)} \right) - 1\over T_{2, \rm q}^{\,(1)} }
\eeq
in terms of the completely known $D$-dimensional one-loop quantities
(see, e.g., Ref.~\cite{AEM79}) with
\bea
\label{Tphiq1}
  T_{\phi, \rm q}^{\,(1)} &\! = \!& 
  - 2\, \cf\: {1 \over \ep}\: \x1^{-\ep}
 \;\;\eqmel\;\; - {2\, \cf \over N}\: {1 \over \ep}\:  \exp\, (\ep \ln N)
\:\: , \\[1mm]
\label{T2q1}
  T_{2, \rm q}^{\,(1)} &\! = \!& 
  - 4\: \cf {1 \over \ep}\: \x1_+^{-1-\ep} + \,\mbox{virtual} 
 \;\;\eqmel\;\; 4\, \cf \:{1 \over \ep^2}\: ( \exp\, (\ep \ln N) - 1 )
\eea
at leading-logarithmic accuracy in both $x$- and $N$-space.
Together with Eq.~(\ref{TijLL}) above, these three relations completely specify
the LL contributions to $T_{\phi, \rm q}$ and $T_{2, \rm g}$ to all orders in 
$\as$ and $\ep$.

As the leading-log expression for $T_{\phi, \rm g}$ are completely analogous
to Eqs.~(\ref{T2qn}), (\ref{C2res}) and (\ref{T2q1}) for $T_{2, \rm q}$, we are
now ready to perform the all-order mass factorization of $T_{\phi, \rm q}$ 
and $T_{2, \rm g}$. We carry out this procedure via expanding all relevant 
expressions to a finite, but very high order using {\sc Form} \cite{FORM3},
and finally deduce the all-order leading-logarithmic splitting functions and 
coefficient functions. Starting with $T_{\phi, \rm q}$, the result for the
former reads
\beq
\label{PgqLL}
  P_{\rm gq}^{\:\rm LL}(N,\as) \;\;=\;\;
  {\cf \over N}\: {\as \over 2\pi}\; {\cal B}_{\,0} (\tilde{a}_{\rm s} )
  \: , \quad
   \tilde{a}_{\rm s} \:\:=\:\:
   {\as \over \pi} \: (\cf\! - \:\!\!\ca)  \ln^{\,2} N
\eeq
with 
\beq
\label{B0}
  {\cal B}_{\,0}(x) \;\:=\;\: \sum_{n=0}^\infty \,\frac{B_n}{(n!)^2} \; x^{\,n}
  \;\:=\;\: 1 \,-\: {x \over 2}
  \; - \;\sum_{n=1}^\infty \,\frac{(-1)^n}{(2n!)^2} \; |B_{2n}|\, x^{\,2n}
\;\; .
\eeq
$B_n$ are the Bernoulli numbers in the standard normalization of 
Ref.~\cite{AbrSteg}: $B_{2n+1} =\, 0$ for $n \geq 1$ and
$$
  B_0 \:=\: 1 \:, \;\; 
  B_1 \:=\: -{1 \over 2} \:, \;\;
  B_2 \:=\: {1 \over 6} \:, \;\;
  B_4 \:=\: -{1 \over 30} \:, \;\;
  B_6 \:=\: {1 \over 42} \:, \;\; \ldots , \;\;
  B_{12} \:=\: -{691 \over 2730} \:, \;\;\ldots
\;\; .
$$
The result for the corresponding coefficient function is given by
\beq
\label{CphiqLL}
  C_{\phi,\rm q}^{\:\rm LL}(N,\as)  \;\:=\;\:  {1 \over N}\:
  \sum_{n=1}^\infty \Big({\as \over 2\pi}\Big)^n \ln^{\,2n-1} N 
  \sum_{a=1}^{n} C_{F}^{\,a} \, C_{\!A}^{\,n-a} \,
  \sum_{j=0}^{n} \: {2^{\:\!j-1} B_{\!j} \over (j\,!)^2} \:
  {(-1)^{j+a} \over (n-j)!} 
  \: \Big( \begin{array}{c} \!\! j-1 \!\! \\ 
                            \!\! a-1 \!\! \end{array} \Big) \!
\:\: .
\eeq
The corresponding results for $P_{\rm qg}$ and $C_{2,\rm g}$ can be obtained 
from Eqs.~(\ref{PgqLL}) and (\ref{CphiqLL}) by replacing one power of $\cf$ by 
$\nf\,$, and then interchange $\cf$ and $\ca$.
Consequently $D_0^{\,(3)}=0$ for both off-diagonal splitting functions in 
Eq.~(\ref{Poffd}) is not at all accidental. In fact, the corresponding 
leading-log contributions vanish at all even orders in $\as$. 
Note also that Eq.~(\ref{PgqLL}) confirms the colour structure of 
Eq.~(\ref{Poffd}) to all orders.
We will provide a more transparent form of Eq.~(\ref{CphiqLL}) below.

The function ${\cal B}_{\,0}(x)$ in Eq.~(\ref{B0}) appears to be new -- at 
least it is not too widely known.
Using the relation between the even-$n$ values of the Riemann $\zeta$-function 
and the corresponding Bernoulli numbers \cite{AbrSteg}, it can be rewritten as 
\beq
\label{B0zeta}
  {\cal B}_{\,0}(x) \;\;=\;\; 1 \:-\: {x \over 2} \:-\: 2 \sum_{n=1}^\infty \:
  {(-1)^n \over (2n)!} \,\zeta_{2n}^{} \Big( {x \over 2\pi} \Big)^{\!2n}
\:\: .
\eeq
The expansion coefficients in the sum differ from the Taylor coefficients of 
$\cos (x/(2\pi))$ by the factor $\zeta_{2n}$. 
Hence, due to $\,\zeta_{2n} \ra 1$ for $\,n \ra \infty\,$, the series 
(\ref{B0}) and (\ref{B0zeta}) converge for all values on $x$. The sum entering 
${\cal B}_{\,0}(2\pi i)$ is known \cite{Sloane,A093721}, if not in a closed form
\cite{W-zeta}.

\begin{figure}[tb]
\vspace*{1.4cm}
\hspace*{6.1cm}\BLUE{\large ${\cal B}_{\,0}(x)$}
\vspace*{-2.1cm}

\centerline{\epsfig{file=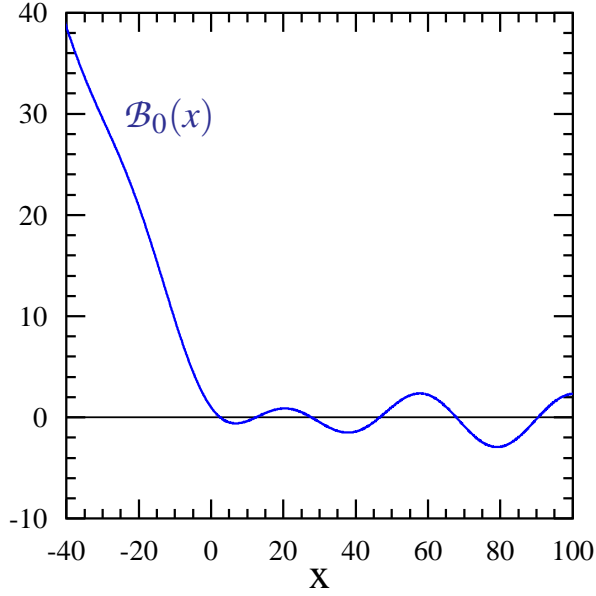,width=8cm,angle=0}}
\vspace{-3mm}
\caption{ \label{fig2}
The function ${\cal B}_{\,0}(x)$ in Eq.~(\ref{B0}), evaluated using its defining 
Taylor expansion.}
\end{figure}
 
\noindent
The numerical behaviour of ${\cal B}_{\,0}(x)$ is illustrated in Fig.~\ref{fig2}.
The even part turns out to oscillate around $|x|/2$, resulting in oscillations 
around $y = 0$ for positive and $y = -x$ for negative values of~$x$, 
respectively.
Related functions, which we expect to enter a generalization of the present
resummation to next-to-leading large-$N$ accuracy, are given by
\beq
  {\cal B}_{\,1}(x) \;=\;
  \sum_{n=0}^\infty \:\frac{B_n}{n!(n+1)!} \; x^{\,n}
\:\: , \quad
  {\cal B}_{-1}(x) \;=\;
  \sum_{n=1}^\infty \:\frac{B_n}{n!(n-1)!} \; x^{\,n}
\:\: .
\eeq
These functions are related to ${\cal B}_{\,0}(x)$ by
\beq 
  \frac{d}{dx}\: (x {\cal B}_{\,1}) \;=\; {\cal B}_{\,0} 
\:\: , \quad
  \frac{d}{dx}\: {\cal B}_{\,0} \;=\; {1 \over x}\: {\cal B}_{-1}
\:\: .
\eeq

Having determined the leading-log off-diagonal splitting functions and
coefficient functions to all orders in $\as$, we are now in a position to prove
(or disprove) the LL part of the conjecture \cite{SMVV1} of the vanishing 
double-logarithmic contributions to the singlet physical kernel for the 
structure functions $F_2$ and $F_\phi$. At this accuracy $\beta(\ar)$ can be 
neglected in Eq.~(\ref{Kdef}) above, leaving $K \,\cong\, C P\, C^{-1}$ with
\beq
\label{KLLprep}
  C^{\,-1} \:\cong\: {1 \over  C_{2,\rm q}^{}\, C_{\phi,\rm g} } \,
  \left( \! \begin{array}{rr} C_{\phi,\rm g}^{}\! & \! -C_{2,\rm g}^{}\! 
  \\ -C_{\phi,q}\! & C_{2,q}^{}\! \end{array}\! \right)
\:\: , \quad
  P^{\,(n\geq 1)} \:\cong\: 
  \left( \! \begin{array}{cc}     0       & \!P_{\rm qg}^{\,(n)}\! \\
                            P_{\rm gq}^{\,(n)}\! &     0   \end{array}\! \right)
\:\: ,
\eeq
recall the notational convention below Eq.~(\ref{ZiiLL}), which yields
\bea
\label{Kp2res}
  K_{\phi 2} & \!\!\cong\!\! & 
     ( C_{2,\rm q}^{} )^{-1} \left\{
     C_{\phi,\rm g}\, P_{\rm gq} \,+\, C_{\phi,\rm q}\, \as 
        \left( P_{\rm qq}^{\,(0)} - P_{\rm gg}^{\,(0)} \right) \right\}
\:\: , \\[1mm]
\label{K2pres}
  K_{2\phi } & \!\!\cong\!\! &
    ( C_{\phi,\rm g} )^{-1} \left\{ 
     C_{2,\rm q}^{}\, P_{\rm qg\,} \,+\, C_{2,\rm g}^{}\, \as 
         \left( P_{\rm gg}^{\,(0)} - P_{\rm qq}^{\,(0)} \right) \right\} 
\:\: .
\eea
Inserting Eq.~(\ref{C2res}) and our results (\ref{PgqLL}) and (\ref{CphiqLL})
into Eq.~(\ref{Kp2res}), and the corresponding relations into 
Eq.~(\ref{K2pres}), the right-hand sides are indeed found to vanish at all
orders $\as^{\,n \geq 2}$.

As all quantities entering Eq.~(\ref{Kp2res}) and Eq.~(\ref{K2pres}) are known
in closed forms (among which we now include ${\cal B}_{\,0}$), these relations
can now be used to cast Eq.~(\ref{CphiqLL}) into the more transparent form
\beq
\label{CphiqLL2} 
   C_{\phi,\rm q}^{\:\rm LL}(N,\as)  \;\:=\;\: 
  {1 \over 2N \ln N} \:{\cf \over \cf - \ca}
   \left\{ \exp \,(2\,\ca \ar \ln^{\,2} N) \, {\cal B}_{\,0} (\tilde{a}_{\rm s})
   - \exp \,(2\, \cf \ar \ln^{\,2} N) \right\}
\:\: ,
\eeq
where the two exponentials are the LL threshold expressions for $C_{\phi,\rm g}$
and $C_{2,\rm q}$ \cite{C2LL}, and $\ar$ and $\,\tilde{a}_{\rm s}$ have been
defined in Eqs.~(\ref{Pexp}) and (\ref{PgqLL}). The corresponding result for 
$C_{2,\rm g}^{\:\rm LL\,}$ is obtained from Eq.~(\ref{CphiqLL2}) by the 
colour-factor replacement given below Eq.~(\ref{CphiqLL}) which includes 
$\tilde{a}_{\rm s} \ra - \,\tilde{a}_{\rm s}$.
Unlike the LL splitting functions, the coefficient functions do not vanish for
for $\cf = \ca$ -- but the curly bracket in Eq.~(\ref{CphiqLL2}) does, cancelling
the corresponding pole in the prefactor.

To summarize, we have derived all-order expressions for the 
large-$x/\,$large-$N$ leading logarithmic (LL) contributions to the 
off-diagonal splitting functions $P_{\rm gq}$ and $P_{\rm qg}$ and the 
corresponding coefficient functions $C_{\phi,\rm q}$ and $C_{2,\rm g}$ in 
Higgs- and gauge-boson exchange deep-inelastic scattering.
Our results show that the LL coefficient for the former two quantities vanish 
at all even orders in the strong coupling $\as$ and confirm that, as 
conjectured in Ref.~\cite{SMVV1}, the leading double-logarithmic contributions 
to the flavour-singlet physical evolution kernels $K_{\phi 2}$ and $K_{2 \phi}$
vanish at all orders.
The key relation have been written down in $N$-space in Eqs.~(\ref{PgqLL}) and 
(\ref{CphiqLL2}), but can be readily inverted back to $x$-space using the second
part of Eq.~(\ref{Logtrf}), in that latter case using the series form 
(\ref{CphiqLL}).
 
The above results for the LL perturbative functions entering DIS (with space-%
like $q^2 \equiv - \Qs$) can be carried over directly the time-like domain 
of semi-inclusive electron-positron annihilation or $Z$ and Higgs-boson decay, 
which is related to the former case by a suitably defined (but for the present
LL contribution essentially trivial) analytic continuation, see, e.g., 
Refs.~\cite{CFP80,PTas3}.

There is scope for improving upon the rigour of the present derivation of the 
crucial relations  (\ref{TijLL}) and (\ref{Tphiqn}) in the future. On may 
expect such as improvement, and other interesting results, from the application
of alternative approaches to deep-inelastic scattering, such as soft-collinear 
effective theory (SCET) \cite{SCET} or the recent path-integral formulation for 
(sub-)$\,$leading threshold contributions \cite{LSW08}.
Within the present framework an extension to at least the next-to-leading
logarithms definitely appears feasible, and we plan to report on this issue in a
later publication.

\vspace*{5mm}
\noindent{\bf Acknowledgments}
 
\noindent
I am grateful to S. Moch for pointing out Ref.~\cite{Sloane} to me and for 
critically reading the manuscript. It is a also pleasure to thank 
J. Vermaseren for helpful comments concerning the effective high-order 
evaluation of nested expansions in {\sc Form}.
This research has been supported by the UK Science \& Technology Facilities 
Council (STFC) under grant numbers PP/E007414/1 and ST/G00062X/1.
 
{\footnotesize
\setlength{\baselineskip}{0.5cm}

}


\begin{thebibliography}{10}
\bibitem{W-zeta} See, e.g., mathworld.wolfram.com/RiemannZetaFunction.html,
where this sum is denoted as $C_3$
\end{thebibliography}
\end{document}